\documentclass[9pt,twocolumn,twoside]{osajnl}

\journal{pr} 

\setboolean{shortarticle}{False}
\usepackage{stfloats}

\title{Fast and wide-band tuning single-mode microlaser based on fiber Fabry-P\'erot microcavities}

\author[1,2]{Xin-Xia Gao}
\author[1,2,*]{Jin-Ming Cui}
\author[1,2]{Zhi-Hao Hu}
\author[1,2]{Chun-Hua Dong}
\author[1,2]{Jian Wang}
\author[1,2]{Yun-Feng Huang}
\author[1,2,$\dagger$]{Chuan-Feng Li}
\author[1,2]{Guang-Can Guo}

\affil[1]{CAS Key Laboratory of Quantum Information, University of Science and Technology of China, Hefei, 230026, China,}
\affil[2]{CAS Center For Excellence in Quantum Information and Quantum Physics, University of Science and Technology of China, Hefei, 230026, People's Republic of China.}

\affil[*]{Corresponding author: jmcui@ustc.edu.cn}
\affil[$\dagger$]{Corresponding author: cfli@ustc.edu.cn}

\begin{abstract}
A narrow linewidth laser operating at the telecommunications band combined with both fast and wide-band tuning features will have promising applications. Here, we demonstrate a single-mode (both transverse and longitude mode) continuous microlaser around 1535 nm based on a fiber Fabry-P\'erot microcavity, which achieves wide-band tuning without mode hopping to 1.3 THz range and fast tuning rate to 60 kHz, yields a frequency scan rate of $1.6\times10^{17}$ Hz/s. Moreover, the linewidth of the laser is measured as narrow as 3.1 MHz. As the microlaser combines all these features into one fiber component, it can serve as the seed laser for versatile applications in optical communication, sensing, frequency-modulated continuous-wave radar and high resolution imaging.

\end{abstract}

\setboolean{displaycopyright}{true}

\makeatother

\usepackage{babel}
\begin{document}
\maketitle

\section{Introduction}

Lasers with properties of low threshold, single-mode output, narrow
linewidth, fast tuning rate and wide tuning range are desired in many
studies \citep{sun2010stable,wu2015monolayer,smith1991narrow,huang2008nanoelectromechanical},
combing these features into one device is critical and feasible for
many practical applications. For this purpose, microlasers supported
by optical microcavities with high-quality factor (Q) are ideal platforms,
as the microcavity have a small mode volume (V) and a large Q/V value,
which implies the ability to realize low threshold and narrow linewidth
lasers \citep{he2013whispering}. In previous research, microlasers
are mainly based on whispering gallery mode (WGM) microcavities, which
have made great progress in the past decades. \citep{spillane2002ultralow,he2013whispering,zhu2018all,yang2017tunable,yang2003gain}.
However, it's still challenging to implement all of the above features
on one device. Although a lot of tunable lasers are proposed in WGM
resonators \citep{yang2017tunable,zhu2018all,ostby2007ultralow,ward2016glass,liu2014all},
tunable microlasers in nm range without mode hopping have not been
reported yet. The upcoming fiber Fabry-P\'erot cavities (FFPCs) with
concave mirrors are widely researched in recent years \citep{hunger2010fiber},
which also provide the high Q factor as that of WGM cavities \citep{hunger2010fiber,janitz2015fabry},
and they have been used in various fundamental and applied research
areas \citep{steiner2013single,janitz2015fabry,zhou2017ultraviolet,hunger2010fiber,haas2014entangled}.
Comparing with the mechanical tuning method of WGM microcavities \citep{dinyari_mechanical_2011},
the FFPCs can be electrically tuned with a much higher mechanical
bandwidth \citep{janitz2017high}. As the FFPC can select a peak mode
while suppressing the others, single longitudinal mode is naturally
feasible. Besides, the short cavity length enables a large free spectrum
range. These properties make the FFPCs an outstanding candidate to
build microlasers.

\begin{figure}
\centering %
\fbox{\includegraphics[width=8.5cm]{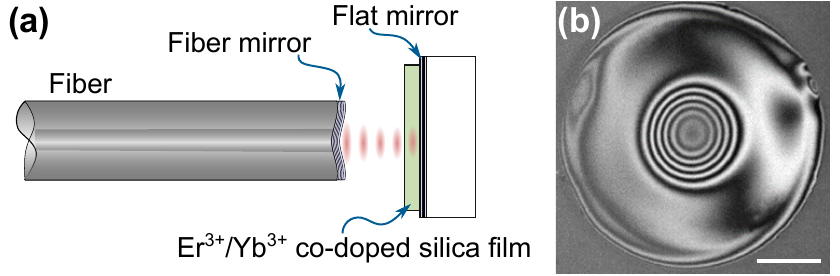}}

\caption{\label{fig:Cavity} Schematic of the FFPC for lasing. (a) The FFPC
is formed by a concave fiber mirror and a flat mirror. An $\mathrm{Er^{3+}/Yb^{3+}}$
co-doped silica film (thickness of 35.1 $\mu$m) is set inside the
cavity, and bonded onto the flat mirror. (b) An interferometric image
of the concave surface of the fiber mirror (with the scale bar of
30 $\mu$m), the radius of curvature of the concave mirror is calculated
as 100 $\mu$m from the concentric fringes.}
\end{figure}

\begin{figure*}
\centering %
\fbox{\includegraphics[width=16cm]{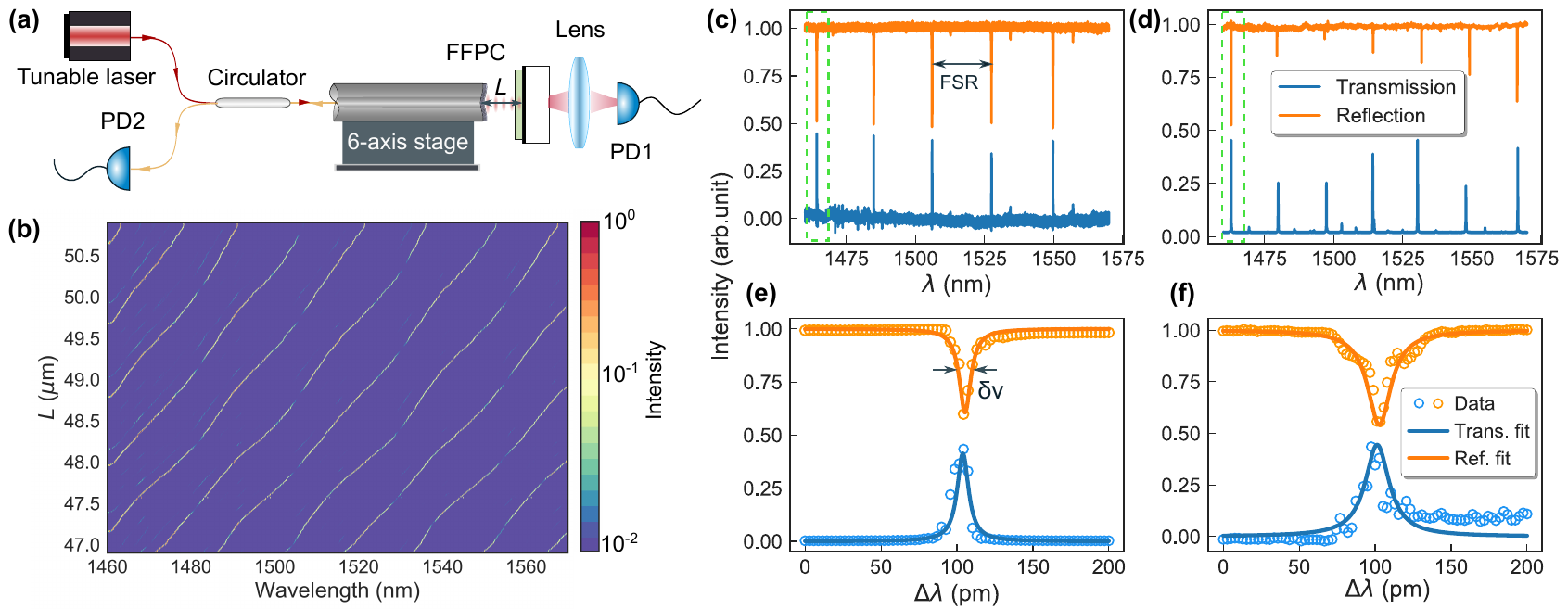}}\caption{\label{fig:cavity spectrum} Measurement of cavity longitude modes.
(a) The scheme to measure the transmission and reflection of the FFPC.
A tunable laser is coupled into the fiber cavity, the transmitted
(reflected) light is detected by PD1 (PD2). A transmission (reflection)
spectrum is obtained by scanning the laser wavelength from 1460 nm
to 1570 nm. (b) A 2-dimensional transmission spectrum by gathering
400 sets of transmission spectrum data as a function of the cavity
length $L$. (c, d) Measured transmission and reflection spectrum
of the bare cavity (film-in-cavity), corresponding $\mathrm{FSR}$
in length to 22.07 nm (17.6 nm), respectively. (e, f) Close-up view
of the peak marked in (c, d), respectively. Lorentzian lineshape is
used to fit the data, the fitted linewidth corresponding $\delta\nu$
in length to 10 pm (17 pm), respectively.}
\end{figure*}

In this work, we demonstrate a microlaser based on an FFPC. The laser
device is a compact module with a single-mode fiber output, which
is capable of emitting single-mode (both transverse and longitude
mode) laser around 1535\ nm. The microlaser wavelength can be electrically
tuned in 10 nm (1.3 THz) range without mode hopping, fast tuning rate
is also realized on our microlasers as the FFPC device is designed
with high mechanical bandwidth \citep{janitz2017high}, and the tuning
bandwidth is tested as 60 kHz, which yields a frequency scan rate
of $1.6\times10^{17}$ Hz/s. The linewidth of the laser is measured
as 3.1 MHz, corresponding a coherence length of 66 m. In general,
we present that a wide-band tunable laser without mode hopping has
the characteristics of fast tuning speed, single fundamental mode,
and narrow linewidth simultaneously.

Benefit from the combined qualities of the device, it has great potential
applications in optical communication, sensing, frequency-modulated
continuous-wave (FMCW) radar \citep{numata2012precision,roos2009ultrabroadband,zhang2019laser},
and high resolution imaging. The range resolution $\delta z$ of an
FMCW measurement is determined by $\delta z=c/2B$ \citep{Burdic1968,zheng2004analysis},
where $c$ is the speed of light, $B$ is the total frequency excursion
of the source. Therefore, the spatial resolution of an imaging system
is inversely proportional to the chirp bandwidth, and the longest
range of the distance measurement is limited by the coherence length
which is determined by the linewidth of the laser. Besides, the laser
seed with high chirp rate can significantly suppress the simultaneous
stimulated Brillouin scattering (SBS) that is currently limiting the
output power of narrow-linewidth fiber amplifiers \citep{white20171,white2012suppression}.
It also can serve as optical beat sources of continuously tunable
terahertz (THz) radiation. \citep{jeon2011rapidly,mittleman2003terahertz}.
Conventionally, distributed-feedback lasers (DFBs), Vertical-Cavity
Surface-Emitting Lasers (VCSELs) and external-cavity diode lasers
(ECDLs) are widely used in those applications \citep{white2012suppression,satyan2010chirp,satyan2009precise}.
However, DFB lasers have a low chirp of $10^{14}$ Hz/s and a limited
tuning range to GHz \citep{satyan2009precise}, and broadband ECDL
to 5 THz has much lower chirp of $6\times10^{12}$ Hz/s \citep{roos2009ultrabroadband}.
Most VCSELs can achieve the linear chirp of $5\times10^{15}$ Hz/s
\citep{white2012suppression}, although a 100$\times$ faster chirp
is obtained by moving the external mirror, the coherent length is
limited to 5 m due to the 40 MHz linewidth \citep{white20171}. Compared
with the traditional swept lasers, our FFPC microlaser demonstrated
the most rapid chirp under THz tuning range with narrow linewidth at the same time. 
\begin{figure*}[t]
\centering %
\fbox{\includegraphics[width=17cm]{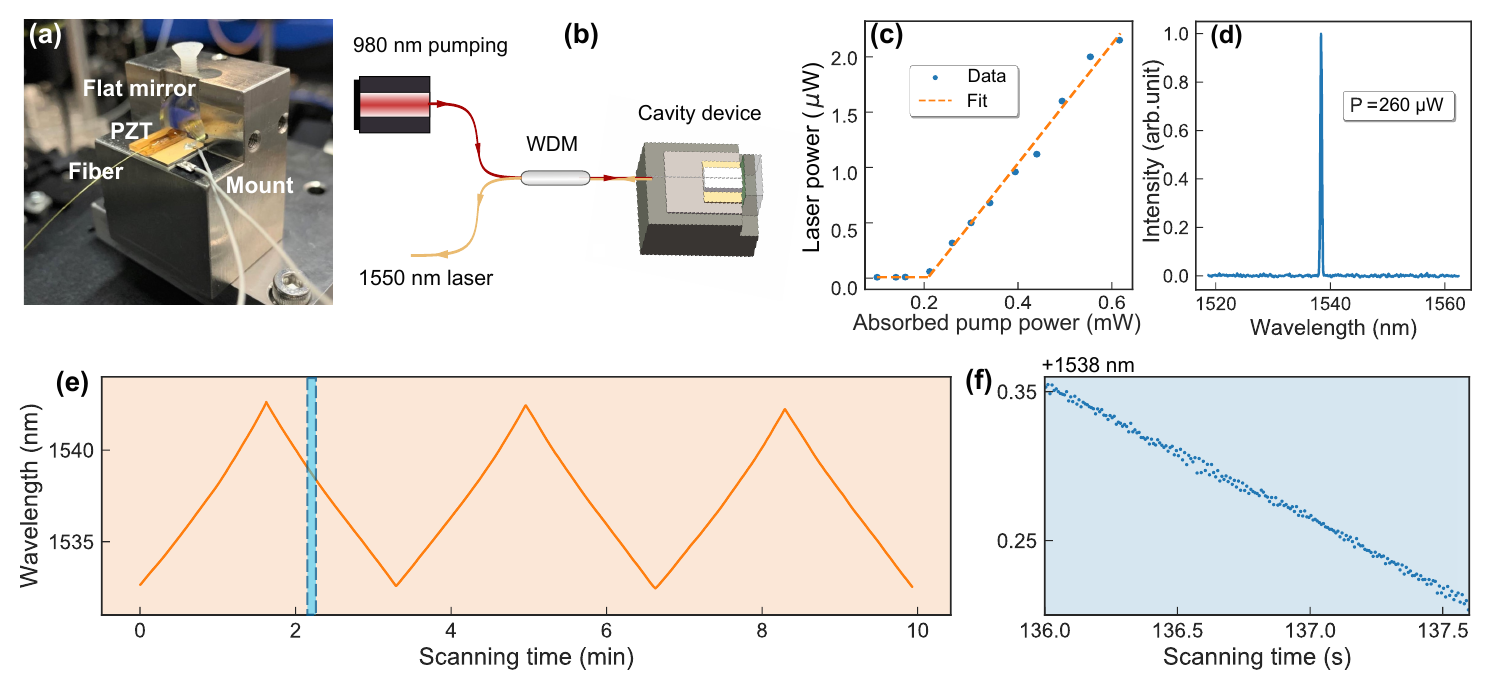}}

\caption{\label{fig:threshold}(a) An image of the assembled microlaser device,
a thick stainless steel bracket is used to support the device, the
fiber mirror is glued on a shear PZT and the flat mirror is mounted
on the right part of the bracket. (b) Schematic of the fiber cavity
laser, a 980 nm laser is used as the pumping source, a fiber-based
WDM (980/1550 nm ) is used to separate the 1550 nm laser output from
the 980 nm pumping input. (c) Laser threshold measurement of the fiber
cavity laser, the threshold is 210 $\mu$W with pump wavelength at
980 nm and lasing wavelength at 1538 nm. (d) The lasing spectrum beyond
the threshold, which indicates a single longitude mode. (e, f) Tunable
range measured by a wavelength meter, (f) is the zoomed chart of the
shade range in (e), which show the laser tunable range without mode
hopping can reach 10 nm.}
\end{figure*}
\begin{figure}[h]
\centering
{\includegraphics[width=8cm]{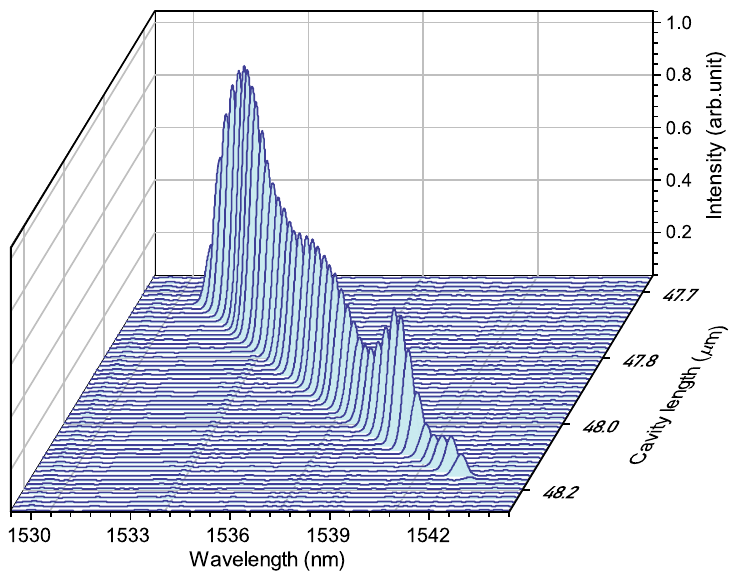}}
\caption{A waterfall diagram of emission spectrum of the cavity laser. It shows the intensity variation versus the tunable wavelength.}
\label{fig:tunable}
\end{figure}
\begin{figure*}[h]
\centering \includegraphics[width=16cm]{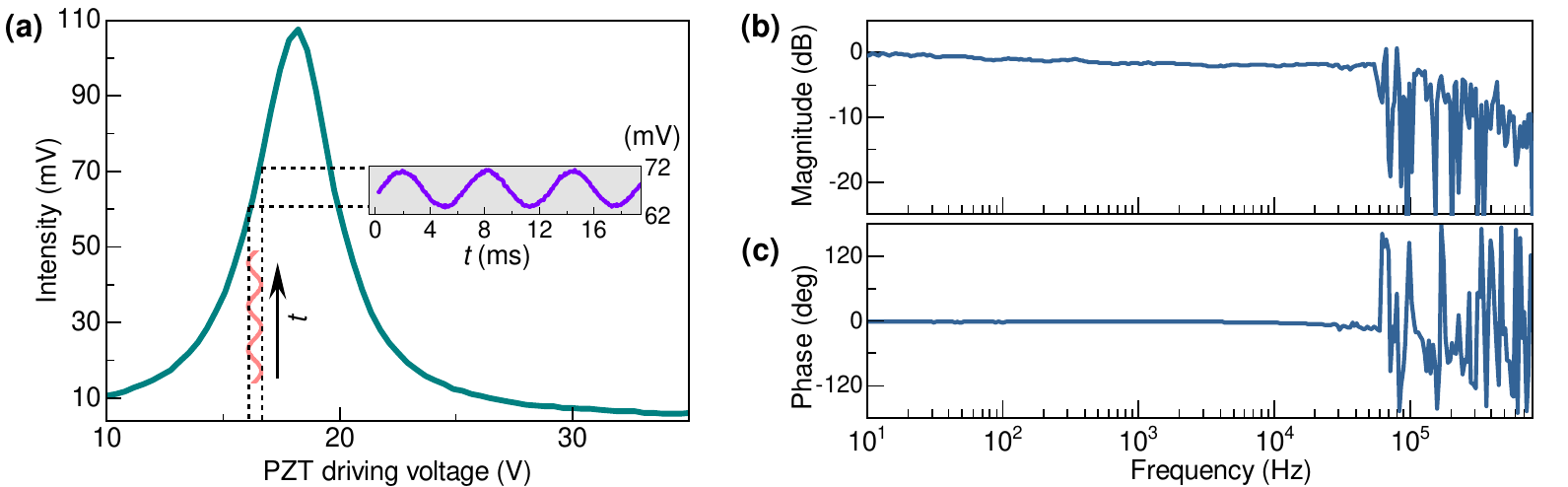}\caption{\label{fig:Resonance}Mechanical bandwidth of the laser device by
measuring the driving response. (a) The envelope of the transmission
spectrum in green and the spectrum response driven by the small sinusoidal
signal. (b) Frequency response diagram of the assembled laser device.
From both magnitude and phase variations, the first direct resonance
is indicated at 60 kHz.}
\end{figure*}

\section{Cavity and cavity measurement}

We use an FFPC as the resonant cavity of a laser, which is formed
by a fiber mirror and a flat mirror (Fig.\,\ref{fig:Cavity}(a)).
The fiber mirror is a concave spherical mirror, which is fabricated
on the end facet of a single-mode optical fiber by using a CO2 laser
ablation process \citep{hunger2010fiber,zhou2017ultraviolet}. After
the laser ablation process, the curvature of the machined fiber end
facet can be measured by a white-light profilometer. An interferometric
image of the end-face image is shown in Fig.\,\ref{fig:Cavity}(b),
where concentric fringes correspond to a radius of the curvature (ROC)
of 100 $\mu$m. The flat mirror is a K9 glass mirror with a diameter
of 12\ mm. Both of the two mirrors are coated with distributed Bragg
reflectors by ion beam sputtering, whose high reflection band is between
1400-1650 nm. The reflectivity of the fiber mirror reaches 99.94\%
while the flat mirror has a lower reflectivity of 99.8\% in order
to collect the transmitted light more efficiently. The mirror coating
is composed of $\mathrm{\mathrm{Ta_{2}O_{5}/SiO_{2}}}$ dielectric
stacks and the final layer is $\mathrm{SiO_{2}}$. An $\mathrm{Er^{3+}/Yb^{3+}}$
co-doped silica film with a thickness of 35.1 $\mu$m is inserted
into the cavity, which acts as the gain medium of the microlaser,
it is bonded to the flat mirror by the chemical reaction between $\mathrm{SiO_{2}}$
and NaOH solution \citep{phelps_strength_2018}. The silica film is
doped with $\mathrm{Yb^{3+}}$ concentration of 19.0 wt\% and $\mathrm{Er^{3+}}$
concentration of 1.0 wt\%, whose gain peak is at 1535 nm.

To assemble the cavity, the flat mirror with the doped silica film
is fixed to a mount, and the fiber is clamped to a 6-axis nanoscale
stage (Thorlabs MAX603D/M). The 6-axis nanoscale stage is mainly used
to enable angular alignment and the cavity length control, which is
precisely adjusted by electrically driving the piezo on the stage.

Once a cavity is made up, the transmission and reflection spectrum
of the cavity is measured to determine the finesse of the cavity.
The measurement diagram is present in Fig.\,\ref{fig:cavity spectrum}(a),
a tunable laser (Toptica CTL 1500) and two photodetectors (PD1, PD2)
are used. The light from the tunable laser is injected into the cavity
through a circulator, then the transmitted light from the cavity is
collected onto PD1 by a lens, and the reflected light by the cavity
passes through the circulator to PD2. By scanning the incident wavelength
from 1460 nm to 1570 nm while monitoring the cavity's transition spectrum
at different cavity lengths, we observe a 2-dimensional (2D) transmission
spectrum shown in Fig.\,\ref{fig:cavity spectrum}(b). The cavity
length is changed by increasing the voltage applied to the nanoscale
stage, with a step length of 10\ nm. The transmission and reflection
spectrum is recorded every step, and 400 sets of spectral data corresponding
to a total variation of 4 $\mu$m in cavity length are recorded.

A one-dimensional model is used to fit the waveform in the 2D transmission
spectrum \citep{janitz2015fabry}, the resonant frequencies $\nu$
are given by the equation below
\begin{equation}
\nu\approx\frac{c}{2\pi\left(L-L_{d}+nL_{d}\right)}\left[\pi m-\left(-1\right)^{m}\arcsin\left(\frac{n-1}{n+1}\sin\theta\right)\right],\label{eq:mode structure}
\end{equation}
where $n$ is the refractive index of the doped silica film, $L$
is the cavity length, $L_{d}$ is the thickness of the doped silica
film, $m$ is an integer, $\theta=\frac{L-L_{d}-nL_{d}}{L-L_{d}+nL_{d}}m\pi$.
Fitting Eq.\,\ref{eq:mode structure} to the 2D cavity transmission
spectrum results in $L_{d}$=35.1 $\mu$m, $L=46\,\mu$m and $n=1.58$.

The waveform also can be explained in an intuition way, if the cavity
is considered as two separate cavities, an air cavity and a film cavity,
they have different resonance frequencies, $\nu_{\mathrm{air}}=mc/2(L-L_{d})$
is linear to the cavity length, while $\nu_{\mathrm{film}}=mc/2nL_{d}$
is a constant, which will offer a vertical line in the 2D spectrum.
However, these modes are actually coupled to one another, which leads
to the bent spectral pattern appeared in the figure.

The finesse of the cavity, which expressed as $\mathfrak{\mathcal{F}}=\mathrm{FSR}/\delta\nu$,
can be obtained from the ratio of the free spectral range (FSR) and
linewidth $\delta\nu$ results. A typical spectrum is shown in Fig.~\ref{fig:cavity spectrum}(c-f).
Fig.~\ref{fig:cavity spectrum}(c-d) show the FSRs of the bare cavity
and film-in-cavity respectively. Fig.~\ref{fig:cavity spectrum}(e-f)
are the corresponding close-up views of the peaks marked in (c-d),
respectively. The measurement data are fitted by a Lorentzian lineshape,
and the linewidth $\delta\nu$ is defined as full-width at half-maximum
(FWHM) of the Lorentzian lineshape. The finesse corresponds to $\mathcal{F_{\mathit{\mathrm{bare}}}}=2207$
and $\mathcal{F_{\mathit{\mathrm{film}}}}=1035$ for the bare cavity
and film-in-cavity. The designed finesse of the bare cavity is 2414,
according to $\mathcal{F}=\pi\sqrt[4]{R_{1}R_{2}}/(1-\sqrt{R_{1}R_{2}})$
\citep{hunger2010fiber}, where $R_{1}$, $R_{2}$ is the reflectivity
of the fiber mirror and flat mirror, respectively. And it is in agreement
with the measured result.
The finesse for the film-in-cavity is significantly
dropped, it is mainly owing to the absorption of a doped silica film
in the cavity, as the measured absorption loss of the silica film
is 0.3\% at the wavelength of 1535 nm.

\section{Cavity laser assembly and measurement}

After the properties of the fiber cavity are characterized, we assemble
the cavity to a device with a shear piezoelectric transducers (PZT)
(Noliac CSAP03) to adjust cavity length. They are glued on a piece
of stainless steel mount with a thick base to make the assembled device
emit stable laser. The photo of the device is shown in Fig.\,\ref{fig:threshold}(a).
From the picture we can see the upright part serves as a mount for
the flat mirror, the mirror is stuffed into half an inch hole and
fixed by a snap ring and a top wire at the same time. The bottom left
part is the base for a stack composed of a ceramics piece, a PZT and
a V-groove, where ceramics piece is used for insulation. They are
glued in place using epoxy (Epotek 301). When all of them are prepared,
the fiber mirror will be aligned to the flat mirror with proper cavity
length by the 6-axis stage and glued in the V-groove by UV curing
adhesive (Epotek H20E). A fiber microcavity device is finally accomplished
after the UV adhesive cured, and used in a microcavity laser experiment.

To make the device emit laser near 1550 nm, a 980 nm pumping laser
is coupled into a cavity device through a wavelength division multiplexer
(WDM), as depicted in Fig.\,\ref{fig:threshold}(b), 1550 nm laser
from the microcavity is coupled back from the input fiber of the device,
and injected to the 1550 nm port of the WDM. Next, the identities
of the fiber cavity laser are measured below.

The laser threshold of the fiber cavity laser is measured by changing
the power of the 980 nm pump laser. For the $\mathrm{Yb^{3+}}$/ $\mathrm{Er^{3+}}$co-doped
silica film, the absorption coefficient of 980 nm laser is measured
as 4.5\%. The output intensity of the microlaser at 1538 nm along
with the absorbed pumping power is plotted in Fig.\,\ref{fig:threshold}(c).
Based on a linear fitting, a threshold of 210 $\mu$W is demonstrated.
Above the threshold, the laser output power increases linearly with
the absorbed pump power. To verify the output laser has a single longitude
mode, the laser beyond the threshold is measured by an optical spectrum
meter (HORIBA iHR 550), the measured spectrum is shown in Fig.\,\ref{fig:threshold}(d).
It is also in single transverse mode because the output laser is coupled
into the fundamental fiber mode.

The wavelength tunability is a very important function for a microlaser.
In this work, we achieve a large tunable range of the output wavelength
from 1532 nm to 1542 nm by electrically controlling the cavity length.
The free stroke of the shear PZT is 1.5 $\mu$m, as a cavity length
variation of 0.5\ $\mu$m (from 47.7 $\mu$m to 48.2 $\mu$m) can
drive the cavity mode from 1532 nm to 1542 nm through Fig.\,\ref{fig:threshold}(e),
the PZT has the ability to tune the laser wavelength more than 10
nm. To measure the no-hopping tunable range, we couple the output
laser into an infrared wavelength meter, and synchronously record
the wavelength with a 5 mHz sawtooth signal driving the PZT. The measurement
result is shown in Fig.\,\ref{fig:threshold}(e), the dynamic curve
of laser wavelength is also a sawtooth wave that is consistent with
the drive signal of the piezo, and shows a peak-to-peak value of 10
nm. To further confirm whether it tunes without mode hopping, we present
a partial enlargement view of Fig.\,\ref{fig:threshold}(e) in Fig.\,\ref{fig:threshold}(f).
It intercepts part of the data within 1.5 s to show the details, which
presents a continuous and linear tunable result.
Except for the measurement by a wavelength meter to show the no-hopping tunable range, the data recorded by the optical spectrum meter is also presented in the form of a waterfall diagram as shown in Fig. \ref{fig:tunable}. By changing the cavity length 12.5 nm each step, 45 sets of laser spectra are grouped together in the waterfall diagram, which shows the intensity variation during wavelength tuning.

Here, we show the wide-band tunable laser can be tuned rapidly at
the same time. Benefit from the high resonant frequency of the shear
PZT and the small mass of the fiber mirror, where the shear PZT is
used to scan the cavity length, the theoretically achievable tunable
frequency is limited by the no-load resonance frequency of the shear
PZT about 1.75 MHz. However, the mechanical resonance frequency of
the laser device may be a more restrictive restriction.
\begin{figure}[t]
\centering \includegraphics[width=8.5cm]{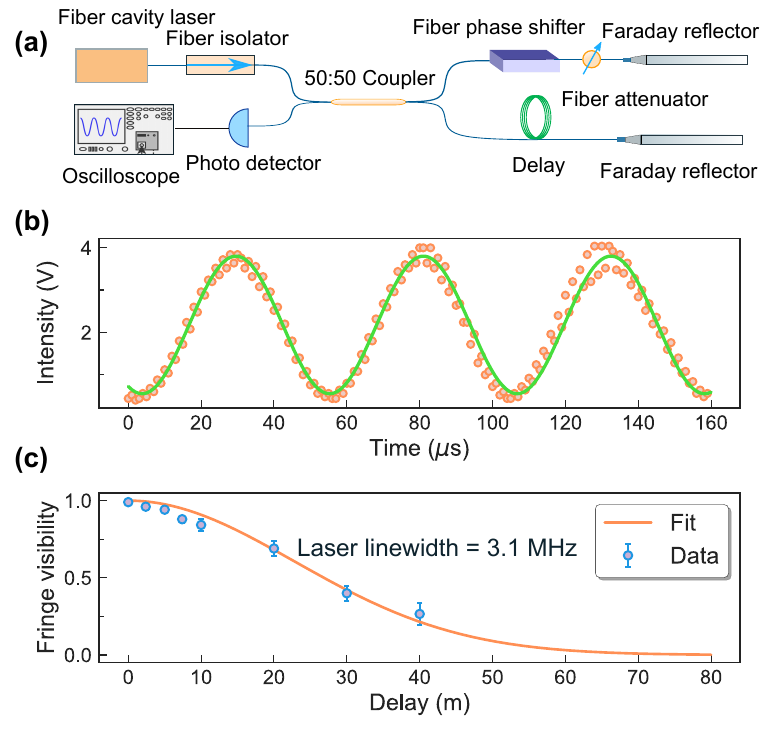}

\caption{\label{fig:Schematic-of-linewidth}Measurement of the laser linewidth.
(a) Schematic of laser linewidth measurement, showing the Michelson
interferometer with two arms: a short arm with a fiber phase shifter,
a long arm producing optical path difference by different-length delay.
Each arm is retro-reflected by the Faraday reflector and a fiber attenuator
to guarantee the same intensity of the two arms. The interference
fringes of the two arms are shown on an oscilloscope. (b) The interfere
fringes at the delay length of 20 m (fit overlaid in green). (c) The
measured results (blue) are fitted by the theoretical curve (orange)
corresponding to the laser linewidth of 3.1 MHz.}
\end{figure}

A frequency response analyzer is used to investigate the mechanical
resonance frequency of the fiber cavity laser assembly. At first,
we measure a transmission spectrum by coupling the Toptica 1550 nm
laser to the cavity and scanning the PZT to change cavity length,
the envelope of one transmission peak is presented in Fig.\,\ref{fig:Resonance}(a).
Then we create a \textquotedblleft small-signal\textquotedblright{}
from the full spectrum by applying a certain bias voltage and small
amplitude sinusoidal modulation to the electrode of the PZT. Driven
by the small sinusoidal signal, the spectrum response with a synchronized
frequency (160 Hz) is shown in the picture. Based on this small signal,
we measure the mechanical resonance frequency by monitoring the magnitude
and phase stability of the response signal in the frequency domain.
The results recorded by a frequency response analyzer is presented
in Fig.\,\ref{fig:Resonance}(b). As the sinusoidal drive frequency
is increased, we eventually encounter mechanical resonances that induce
the first $\pi$-phase delay and violent moves of amplitude at 60
kHz. Therefore, we achieve a fast tuning rate at 60 kHz. And the mechanical
resonance frequency can be further improved by minimizing the length
of overhanging fiber from the piezo, reducing the thickness of the
epoxy holding the fiber and optimizing the geometry of the mount \citep{janitz2017high}.

Besides, we independently demonstrate the laser linewidth by measuring
fringe contrast based on a Michelson interferometer with different-length
delays. Fig.\,\ref{fig:Schematic-of-linewidth}(a) presents the experimental
configuration of the measurement system. A Michelson interferometer
splits the 1550 nm laser into two arms: a short arm with a fiber phase
shifter and a long arm with relatively long round trip time delay
$\Delta t$. As in the standard Michelson configuration, each arm
is retro-reflected by a Faraday reflector, correctly compensating
for polarization changes which are caused by the polarization-free
fiber. Here is also a fiber attenuator to guarantee the equal intensity
of the two arms. Recombination of the two arms generates interference
fringes that can be detected by a photo-diode and presented on an
oscilloscope. Fig.\,\ref{fig:Schematic-of-linewidth}(b) is an interference
fringe when the delay length is at 20 m. Fringe contrast can be calculated
by
\begin{equation}
V=\frac{I_{\mathrm{max}}-I_{\mathrm{min}}}{I_{\mathrm{max}}+I_{\mathrm{min}}},\label{eq:fringe}
\end{equation}
where $I_{\mathrm{max}}$ and $I_{\mathrm{min}}$ are maximum and
minimum intensity in the interference fringe.

On the other hand, considering the power spectrum of the laser with
a Gaussian noise, the fringe contrast $V$ is proportional to the
normalized Fourier transform of the source power spectrum \citep{simon1987fringe},
which is
\begin{equation}
V=\mid\mathrm{gaus}(\Delta\nu\Delta t)\mid,\label{eq:fringe contrast}
\end{equation}
where $\mathrm{gaus}(\nu)=\exp(-\pi\nu^{2})$, $\Delta\nu$ is the
linewidth of the laser, the time difference $\Delta t$ corresponds
to the optical path difference $L=c\Delta t$, which arises from the
delay in the long arm. We set different fiber delays and measured
the corresponding fringe contrasts, the results are shown in Fig.\,\ref{fig:Schematic-of-linewidth}(c)
. By applying Eq.\,\ref{eq:fringe contrast} to fit the measured
results, the laser linewidth is obtained as $3.1\pm0.2$ MHz.

\begin{figure}[htbp]
\centering
{\includegraphics[width=8cm]{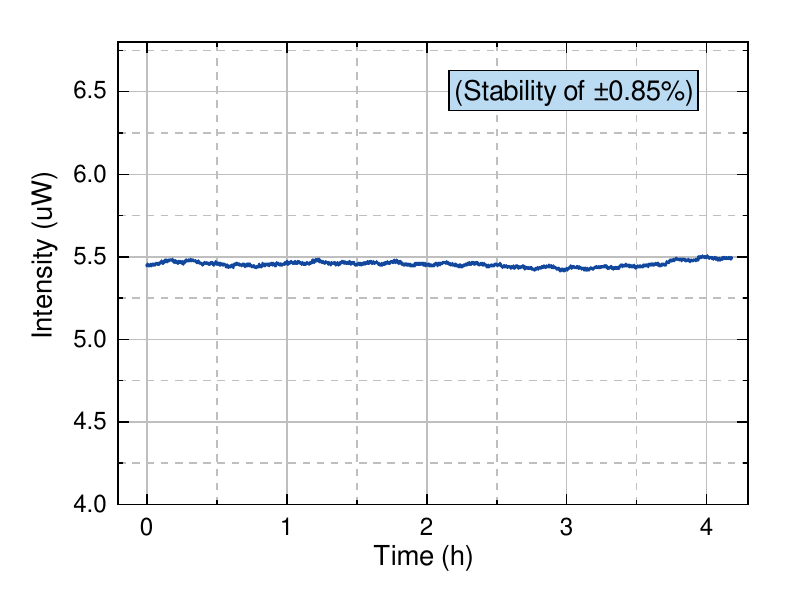}}
\caption{The stability measurement of the laser intensity, corresponding to the stability of $\pm0.85\%$ for 4 hours.}
\label{fig:stability}
\end{figure}

In order to confirm the stability of the output laser, we measure the laser power at a fixed wavelength of 1535 nm. The results are recorded by an optical power meter with a sample interval of 1 second for continual 4 hours. Under ambient conditions, the measured intensity of the output laser is displayed in Fig. \ref{fig:stability}, which corresponds to the stability of $\pm0.85\%$. The stability of the output power can be further improved if a temperature-controlled environment is created.

\section{Conclusion and outlook}

In summary, we have designed and fabricated an infrared single-mode
microlaser based on an FFPC, demonstrated fast chirp of $\ensuremath{1.6\times10^{17}}$
Hz/s, wide-band tuning range of 1.3 THz without mode hopping and narrow
linewidth of 3.1 MHz at the same time. As a bare FFPC with concave
mirrors has achieved high finesse in excess of 100000 \citep{hunger2010fiber},
the finesse of the FFPC in our device can be further improved by decreasing
the absorption of the doped film inside the cavity, thus the linewidth
of the laser can be much narrower than 1 MHz. The integrated design
of the cavity mount allows a high mechanical bandwidth of 60 kHz,
which corresponds a frequency scan rate of $1.6\times10^{17}$ Hz/s.
The tuning frequency can also be further increased by reducing the
load of epoxy on the shear PZT and optimizing the geometry of the
mount. These combined properties make it can enhance the spatial resolution,
the measuring range and the measurement speed in optical communication,
sensing, FMCW radar and high resolution imaging.

\section*{Acknowledgments}

We acknowledge funding support from the National Key Research and
Development Program of China (Nos. 2017YFA0304100, 2016YFA0302700),
National Natural Science Foundation of China (Nos. 11774335, 11734015,
11804330, 11821404), Key Research Program of Frontier Sciences,
CAS (No. QYZDY-SSW-SLH003), Science Fundation of the CAS (No. ZDRW-XH-2019-1),
the Fundamental Research Funds for the Central Universities (Nos.
WK2470000026, WK2470000027, WK2470000028), Anhui Initiative
in Quantum Information Technologies (Nos. AHY020100, AHY070000).

\section*{Disclosures}
\noindent\textbf{Disclosures.} The authors declare no conflicts of interest.

\bibliography{MicroCavityLaser.bib}

\end{document}